\documentclass[12pt]{article}
\usepackage{graphicx}

\date{\today}
\title{Unveiling the Surface Structure of Amorphous Solid Water via Selective Infrared Irradiation of OH Stretching
Modes}

\author{J.~A. Noble\thanks{Laboratoire Physique des Interactions Ioniques et Mol\'{e}culaires, UMR 7345-CNRS, Aix-Marseille Universit\'{e}, Centre St-J\'{e}r\^{o}me, 13397 Marseille Cedex 20, France.}, C. Martin$\rm{^*}$, H.~J. Fraser\thanks{Astronomy Division, CEPSAR, Department of Physical Sciences, The Open University, Walton Hall, Milton Keynes, MK7 6AA, UK.}, P. Roubin$\rm{^*}$ \& S. Coussan$\rm{^*}$}

\begin{document}

\maketitle

\begin{abstract}
In the quest to understand the formation of the building blocks of life, amorphous solid water (ASW) is one of the most widely studied molecular systems. Indeed, ASW is ubiquitous in the cold interstellar medium (ISM), where ASW-coated dust grains provide a catalytic surface for solid phase chemistry, and is believed to be present in the Earth's atmosphere at high altitudes. It has been shown that the ice surface adsorbs small molecules such as CO, N$_2$, or CH$_4$, most likely at OH groups dangling from the surface.
Our study presents completely new insights concerning the behaviour of ASW upon selective infrared (IR) irradiation of its dangling modes. When irradiated, these surface H$_2$O molecules reorganise, predominantly forming a stabilised monomer-like water mode on the ice surface. We show that we systematically provoke ``hole-burning'' effects (or net loss of oscillators) at the wavelength of irradiation and reproduce the same absorbed water monomer on the ASW surface. Our study suggests that all dangling modes share one common channel of vibrational relaxation; the ice remains amorphous but with a reduced range of binding sites, and thus an altered catalytic capacity.\\
\end{abstract}

\begin{figure}
\includegraphics[width=5cm]{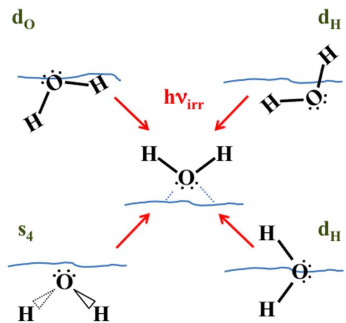}
\caption{Table of Contents graphic}
\end{figure}

\begin{figure}
\includegraphics[width=0.5\textwidth]{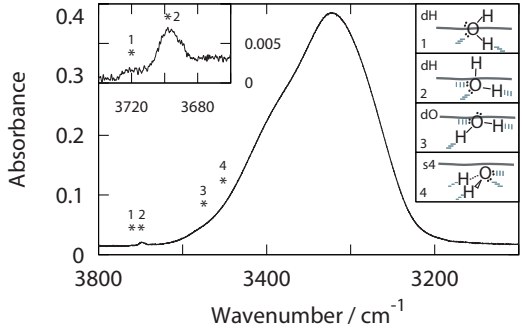}
\caption{The four surface modes of water ice -- dH (3720 and 3698~cm$^{-1}$), dO (3549~cm$^{-1}$), and s4 (3503~cm$^{-1}$) -- are illustrated on a representative ASW sample.}\label{fig:cartoon}
\end{figure}

ASW is a molecular system which has long provoked interest due, in part, to its role in the formation of molecules key to the origins of life\cite{Buch91,Devlin95,Buch97,Manca01A,Manca02C}.
ASW has long been known to accrete small molecules such as CO, H$_2$O, N$_2$, or CH$_4$\cite{Devlin95,Manca00,Collings04}, initiating chemical and photochemical surface reactivity\cite{Devlin95,Watanabe02,Rowland91}. In the ISM, water in the form of ASW is the most abundant solid phase molecular species\cite{Leger79,Whittet83}. The production of molecules, from the most simple, H$_2$\cite{Hollenbach71,Manico01}, to the more complex CH$_3$OH\cite{Watanabe02}, and even precursors to the simplest amino acid, glycine\cite{Danger12}, is catalysed by the ASW surface\cite{Williams96}; both the outer surface and surfaces within its porous structure are involved. The selective IR irradiation of crystalline ice\cite{Focsa03,Focsa06} and water clusters\cite{Buck98} has already been studied. In the former case, the desorption of H$_2$O molecules, and in the latter, the dissociation of clusters, was stimulated.
We are interested in the behaviour of ASW upon selective IR irradiation and have studied the irradiation of the four surface modes of this ice, assigned in the literature\cite{Buch91,Devlin95} and illustrated in Figure~\ref{fig:cartoon}. Theoretical calculations, supported by experimental studies, suggest that water molecules in the dH mode are bi- or tri-coordinated, presenting one free OH bond dangling at the surface; dO molecules present a free oxygen electronic doublet; and s4 molecules have a tetrahedral structure at the surface, which is distorted compared to the tetrahedra of bulk ASW.

In these experiments we prepared a pure ASW sample (Figure~\ref{fig:cartoon}) as follows: deionised water was subjected to multiple freeze-pump-thaw cycles under vacuum to remove dissolved gases. Mixtures of purified H$_2$O and helium (Air Liquide, $\geq$~99.9999~\%) gas were prepared in a stainless steel dosing line with base pressure 10$^{-4}$~mbar in ratios of $\rm{H_2O:He}=1:25$. Ices were produced by depositing the gas mixture directly onto a gold-plated copper surface held at 50~K (to avoid trapping of the vector gas or nitrogen) then cooled to 3.7~K (the cooling typically takes around five minutes due to the high cryogenic power of 0.5~W at 4~K). The cooled surface is located in a high vacuum chamber with a base pressure of 10$^{-8}$~mbar at 3.7~K. IR spectra were recorded in reflection mode using a Bruker 66/S FTIR spectrometer equipped with a MCT detector (4000 -- 800~cm$^{-1}$). Full details of the experimental setup are given in Coussan \textit{et al.}\cite{Coussan12}.  The ices grown in our study were characterised as purely amorphous in nature due to the position of the bulk OH stretch and the characteristic dangling modes at 3720 and 3698~cm$^{-1}$ (see Figure~\ref{fig:cartoon})\cite{Buch91,Devlin95}. Some previous studies using a helium carrier gas have reported the formation of ice nanocrystals or clusters\cite{Rowland91b}, but this can be ruled out due to the absence of an absorption feature centred at 3692~cm$^{-1}$. This band was never observed in any deposition of a H$_2$O:He mixture during the work detailed here. After deposition, ices were selectively irradiated using a tunable IR OPO Laserspec (1.5 -- 4~$\mu$m), pumped at 10~Hz by a pulsed Nd:YAG Quantel Brilliant B laser (1064~nm, pulse duration 6~ns). The average laser power is $\approx$~35~mW in the $\rm{\nu_{OH}}$ domain, except in the range 3520 -- 3500 cm$^{-1}$ where it is $\approx$~10~mW, with a FWHM $\geq$~1.5~cm$^{-1}$. Each irradiation was performed for one hour to ensure saturation of the effects.

\begin{figure}
\includegraphics[width=0.5\textwidth]{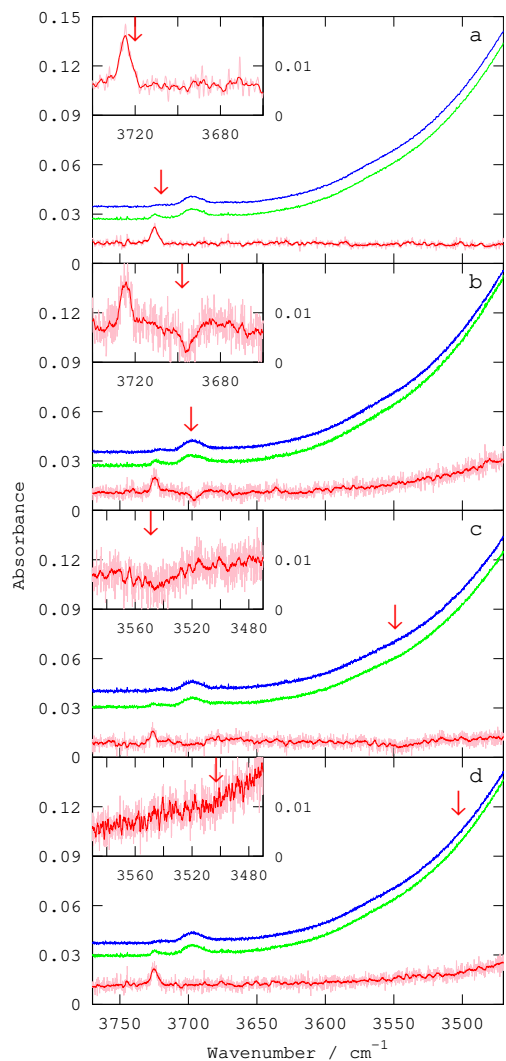}%
\caption{Effect of selective irradiation. Panels show the effects of selective irradiation at (a) 3720 cm$^{-1}$, (b) 3698 cm$^{-1}$, (c) 3549 cm$^{-1}$, and (d) 3503 cm$^{-1}$. The spectra are coloured as follows: before irradiation, blue; after irradiation, green; difference spectrum, red. The insets magnify the irradiated range.}\label{fig:irr}
\end{figure}

\begin{figure}
\includegraphics[width=0.5\textwidth]{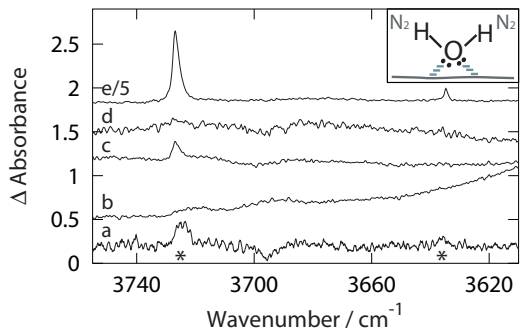}
\caption{ Background deposition experiments. We present the difference spectra of: (a) ASW irradiated at 3698~cm$^{-1}$, reproduced from Figure~\ref{fig:irr}; (b) An ASW sample plus background-deposited water; (c) Spectrum~b plus background-deposited nitrogen; (d) An ASW sample plus background-deposited nitrogen; (e) Spectrum~d plus background-deposited H$_2$O:N$_2$ mixture.}\label{fig:depot}
\end{figure}

Figure~\ref{fig:irr} shows the results of irradiations carried out on the four dangling mode bands. The irradiations provoke permanent ``hole-burnings'' in the irradiated band which are slightly shifted (by up to 3~cm$^{-1}$) compared to the irradiation frequency. The ``hole-burning'' is clearest for the irradiation of dH at 3698~cm$^{-1}$ but is also easily visible for the irradiations at 3720 and 3549~cm$^{-1}$.
Interestingly, upon each irradiation, we observe the growth of a new band centred at $\sim$~3725~cm$^{-1}$, with FWHM $\approx$ 5~cm$^{-1}$.
Both the shifts in frequency upon irradiation, and the narrowness of the ``hole-burning'' effects and the newly created bands at 3725~cm$^{-1}$ illustrate the inhomogeneity of the bands; each band contains a distribution of oscillators, but only one class of oscillator isomerises upon irradiation at a given frequency, producing one new oscillator class. After irradiation at 3698~cm$^{-1}$, a second, smaller peak at 3638~cm$^{-1}$ is also observed (see Figure~\ref{fig:depot}a). Unirradiated ASW samples and their IR spectra were stable and remained unchanged over the timescale of an irradiation study.

The common feature of all the irradiations in Figure~\ref{fig:irr} is the growth of the 3725~cm$^{-1}$ band. What is the source of this new band, previously unidentified in ASW spectra?
Considering energetics only, irradiating between 3720 and 3503~cm$^{-1}$ could potentially break one or two hydrogen bonds, as the average weak H-bond strength is around 1800~cm$^{-1}$. Is the band, therefore, due to H$_2$O surface molecules in a different conformation than those in dH, dO and s4 modes, or is it due to water molecules which have desorbed then re-adsorbed at the surface? The latter response can be immediately discarded based on the results of molecular adsorption studies which have shown a red-shift of the OH dangling frequency upon adsorption of multiple molecular species at the dangling modes\cite{Devlin95,Manca01A,Manca03}. For example, in the case of nitrogen adsorption onto ASW, Manca \textit{et al.}\cite{Manca03} observed a red shift of 22~cm$^{-1}$ of the dH mode\cite{Sadlej95,Buch04}.
In our experiments, the new band at 3725~cm$^{-1}$ is blue-shifted with respect to the dH modes. Moreover, the dynamic vacuum of 10$^{-8}$~mbar rapidly evacuates any desorbing molecules, such that the residual pressure is too low to allow redeposition. Thermal effects are discounted based upon annealing of ASW samples, which provoked a global decrease of the dangling bonds and no production of narrow peaks at 3725~cm$^{-1}$, in agreement with previous studies\cite{Devlin95}. Upon irradiation, the newly produced band is narrow, indicating a single, homogeneous vibrational mode.
Thus, the 3725~cm$^{-1}$ band is clearly not due to a perturbed dH mode, but it rather has dangling OH character, and we propose that it is samples a bi-coordinated H$_2$O molecule, with two dangling OH and two co-ordinated electron pairs (d2H). This structure explains the blue shift of the peak with respect to the dH mode at 3720~cm$^{-1}$ as, because neither of the two OH oscillators is directly hydrogen bonded, the free OH oscillators are less perturbed than the surface dH modes.

We performed further experiments to verify the source of the two new peaks at 3725 and 3638~cm$^{-1}$, as illustrated in Figure~\ref{fig:depot}. The background deposition of water on a pure ASW sample does not result in a new band, but rather the global growth of the dH and bulk OH bands (Figure~\ref{fig:depot}b). However, subsequent background deposition of pure nitrogen,  a molecule present as a low-level source of pollution in the chamber, provokes the appearance of a band at 3726~cm$^{-1}$ (Figure~\ref{fig:depot}c), revealed thanks to the magnifying effect of nitrogen\cite{Hujo11}. 
The background deposition of N$_2$ upon a new ASW sample does not produce a clear peak (Figure~\ref{fig:depot}d), but the background deposition of a H$_2$O:N$_2$ (1:10) mixture provokes the appearance of two narrow bands at 3725~cm$^{-1}$ and 3638 cm$^{-1}$ (Figure~\ref{fig:depot}), with a peak area ratio of approximately 10:1, compared to 6:1 seen after irradiation at 3698~cm$^{-1}$. 
When this deposition spectrum is compared to the spectrum of a gas-phase water monomer (see Table~\ref{table:frequencies}), the two peaks show red shifts of nearly 200~cm$^{-1}$ with respect to the $\rm\nu_3/\nu_1$ modes of the monomer (one always compares gas-phase experimental results with those of a ``classical'' theoretical calculation, as classical calculations are carried out for isolated species. It also allows us to approximate the perturbation induced by the medium). These red shifts  are almost precisely those observed in the solid phase for water in a nitrogen matrix, where the $\rm\nu_3 : \nu_1$ ratio is 8:1, and a H\rm$_2$O-N\rm$_2$ complex.\cite{Coussan98,Coussan06}
Considering our experimental results and the literature, the band at 3725~cm$^{-1}$ is positively attributed to a water monomer interacting with the surface via its two electronic doublets; its large intensity compared to the other dH bands is explained by the magnifying effect of nitrogen, as extensively investigated by Hujo \textit{et al.}\cite{Hujo11}. We suggest that nitrogen molecules, present as a low-level pollutant in the chamber, serendipitously complex the water molecule, as illustrated in Figure~\ref{fig:depot}, stabilising the molecule, preventing any further adsorption, and magnifying the OH stretching bands.
 However, the nitrogen must only be present as a trace pollutant on the surface, as no shifting of the dH peak positions of the ASW is seen, either during cooling of the water ice sample from 50~K to 3.7~K, or during the irradiation period. If nitrogen were present at multilayer concentrations, we would expect to see a red shift\cite{Manca03,Sadlej95,Buch04} of up to 22~cm$^{-1}$.
The $\nu_1$ peak at 3638~cm$^{-1}$ is not observed after irradiation of the other three surface modes, but this is likely due to its low intensity compared to that of $\nu_3$.

\begin{table}[h!]
\caption{Comparison between calculated and observed $\nu_3$ and $\nu_1$ water monomer frequencies. Frequencies are given in cm$^{-1}$.}\label{table:frequencies}
\begin{tabular}{cccccc}

Vibrational & Calculated\cite{Coussan98} & \multicolumn{3}{c}{Observed}                                                                        & Present  \\
mode         &  (gas phase)      & gas phase\cite{Benedict56} & N$_2$ matrix\cite{Coussan06} & H\rm$_2$O-N\rm$_2$ complex\cite{Coussan98} & work \\
\hline
\rm$\nu_3$ & 3924 & 3943 & 3728 & 3730 & 3725 \\
\rm$\nu_1$ & 3822 & 3832 & 3635 & 3640 & 3638 \\
\rm$\Delta_\nu=\nu_3-\nu_1$ & 102 & 111 & 93 & 90 & 87

\end{tabular}

\end{table}

One explanation for the observed ``hole-burning'' is that amorphous ice is unable to relax all of the vibrational energy injected at the surface through bulk relaxation channels. As a result, some fraction of this energy is accumulated at the surface and in the immediate sub-layers, where it induces reconstruction of the surface. We observe saturation of the  ``hole-burning'' events within the timescale of the irradiations performed, suggesting that the rearrangement of surface molecules is not an efficient relaxation channel, but is a minority effect compared to the main relaxation channels via the bulk ice. Thus, the production of the 3725~cm$^{-1}$ band is not the major result of irradiation, but is due to the inability of the ice to fully dissipate the injected energy. It has previously been suggested that ASW is a disordered material which has no long-range organisation\cite{Buch91}, which could help to explain the lack of efficiency in the bulk relaxation channels. Although it is possible that some H$_2$O molecules desorb upon irradiation, as in the studies of Focsa \textit{et al.}\cite{Focsa03,Focsa06}, we consider this effect to be minor because no increase in pressure is observed in the chamber and the energy injected is not enough to break more than two H-bonds.

It is curious that the absorption band at 3725~cm$^{-1}$ is produced upon irradiation of all dangling bonds. The irradiation effects at 3720 and 3698~cm$^{-1}$, in particular, are very similar, except that the ``hole-burning'' of the doubly-coordinated dH is less pronounced than that of the triply-coordinated dH.
Such molecular rearrangement requires only a reorientation of the water molecule and thus is not ``energy consuming'' compared to the energy injected into the system. These results also provide evidence of a local ordering to the ASW structure. Upon irradiation of the dH modes, both the ``hole-burning'' in absorption bands and the newly produced monomer band at 3725~cm$^{-1}$ are relatively narrow (FWHM $\approx$ 5~cm$^{-1}$). This suggests that each surface molecule is surrounded by a locally ordered oscillator network, resulting reproducibly in the production of one oscillator class upon selective irradiation. If we consider irradiation of the s4 band, centred at 3503~cm$^{-1}$, we were unable to observe a definitive ``hole-burning'' event, likely due to the width of the band. However, we observed an increase at 3725~cm$^{-1}$, as for the dH modes. It is unlikely that the tetra-coordinated s4 molecules are themselves ejected from within the surface layer, as this would be highly energetically unfavourable. However, as we see an increase at 3725~cm$^{-1}$, it is likely that the relaxation channel involves the breaking of H-bonds, with two breaks being enough to ``transform'' a s4 molecule into a monomer-like water molecule. The case of dO is likely intermediate between those of dH and s4. During each of these irradiations, we see no interconversion between the modes, suggesting that there is only a single ``surface'' channel for the release of excess vibrational energy.

In this work we have provided new insights into the bonding and structure of the surface molecules in amorphous solid water. The ensemble of our results show that surface modes, in particular the dH dangling bonds, are sensitive to photo-induced rearrangement due to competition between surface reorganisation and the main relaxation channels in the bulk water ice. Rather than desorbing from the surface, molecules embedded in the surface layer become loosely associated with the surface in the form of monomer-like structures interacting through their two free electron pairs. The fortuitous presence of nitrogen in the chamber both promoted the magnification of the OH stretching mode of the monomer-like molecule and stabilised it on the ice surface. Inducing such conformational changes in an ASW surface potentially alters its physicochemical properties, most notably its catalytic potential.

\section*{Acknowledgements}
J.~A. Noble is a Royal Commission for the Exhibition of 1851 Research Fellow

\section*{Author information}
*Corresponding author : S. Coussan, stephane.coussan@univ-amu.fr

\end{document}